\documentclass[%
 reprint,
 amsmath,amssymb,
 aps,
 pre,
]{revtex4-2}
\usepackage{soul}
\usepackage{graphicx}
\usepackage{dcolumn}
\usepackage{bm}
\usepackage{xcolor}%

\newcommand{\hly}[1]{\colorbox{white}{\parbox{\columnwidth}{#1}}}

\newcommand{\highlightmath}[1]{\colorbox{white}{$\displaystyle #1$}}

\newcommand\tmu{\tilde{\mu}}

\renewcommand\L {\mathcal{L}}

\renewcommand{\vec}[1]{\boldsymbol{#1}}

\newcommand{\grad}{\nabla}

\newcommand{\J}{\mathcal{J}}
\newcommand{\Ju}{\mathcal{J}^{\vec{u}}}

\newcommand{\Juo}{\mathcal{J}^{u}_0}
\newcommand{\Juf}{\mathcal{J}^{u}_f}
\newcommand{\JUo}{\mathcal{J}^{\vec{u}}_0}
\newcommand{\JUf}{\mathcal{J}^{\vec{u}}_f}

\newcommand{\Jwf}{\mathcal{J}^{\omega}_f}

\newcommand{\T}{{\cal T}}

\begin{document}

\preprint{APS/123-QED}

\title{Iterative Methods for Navier--Stokes Inverse Problems}

\author{Liam O'Connor$^{1,2}$, Daniel Lecoanet$^{1,2}$, Evan H. Anders$^{3}$, Kyle C. Augustson$^{1,2}$, Keaton J. Burns$^{4}$, Geoffrey M. Vasil$^{5}$, Jeffrey S. Oishi$^{6}$, Benjamin P. Brown$^{7}$}

\address{$^1$Department of Engineering Sciences and Applied Mathematics, Northwestern University, Evanston, IL, 60208 USA}
\address{$^2$Center for Interdisciplinary Exploration and Research in Astrophysics, Northwestern University, Evanston, IL, 60201 USA}
\address{$^3$Kavli Institute for Theoretical Physics, University of California Santa Barbara, Santa Barbara, CA, 93106 USA}
\address{$^4$Department of Mathematics, Massachusetts Institute of Technology, Cambridge, MA, 02142 USA}
\address{$^5$School of Mathematics, Edinburgh University, EH9 3FD, UK}
\address{$^6$Department of Physics and Astronomy, Bates College, Lewiston, ME, 04240 USA}
\address{$^7$Department of Astrophysical and Planetary Sciences, University of Colorado Boulder, Boulder, CO, 80309 USA}

\begin{abstract}
  Even when the partial differential equation underlying a physical process can be evolved forward in time, the retrospective (backward in time) inverse problem often has its own challenges and applications.
  Direct Adjoint Looping (DAL) is the defacto approach for solving retrospective inverse problems, but it has not been applied to deterministic \colorbox{white}{retrospective} Navier--Stokes inverse problems in 2D or 3D.
  In this paper, we demonstrate that DAL is ill-suited for solving retrospective 2D Navier--Stokes inverse problems. 
  Alongside DAL, we study two other iterative methods: Simple Backward Integration (SBI) and the Quasi-Reversible Method (QRM).
  Our iterative SBI approach is novel while iterative QRM has previously been used.
  Using these three iterative methods, we solve two retrospective inverse problems:
  1D Korteweg--de Vries--Burgers (decaying nonlinear wave) and 2D Navier--Stokes (unstratified Kelvin--Helmholtz vortex).
  In both cases, SBI and QRM reproduce the target final states more accurately and in fewer iterations than DAL.
  We attribute this performance gap to additional terms present in SBI and QRM's respective backward integrations which are absent in DAL.
\end{abstract}

\maketitle

\section{Introduction}\label{secintro}

Consider a retrospective inverse problem where the variable of interest $\overline{X}$ is constrained by a partial differential equation (PDE) involving space $x$ and time $t$:
\begin{align}
  \mathcal{F}[\overline{X}(x, t)] = 0 \label{genintro}.
\end{align}
We aim to determine the unique initial condition $\overline{X}(x, 0)$ corresponding to a given final state $\overline{X}(x, t_f)$ where the final time $t_f$ is known.
These ``retrospective inverse problems'' are often ill-conditioned or ill-posed, even when the forward problem is readily solvable.

Despite potential difficulties, retrospective inverse problems arise in a wide variety of contexts.
Scientists and engineers have documented numerous cases where retrospective inversion is feasible and practical.
\cite{Lukyanenko2021} inverted chemical reaction fronts governed by a 1D reaction-diffusion equation.
\cite{Subramanian2020} inverts tumor growth using biophysically-motivated constraints on the initial condition.
\cite{Kabanikhin2020} inverted supersonic supernova expansions governed by the 1D compressible Navier--Stokes equation.
\cite{Liu2008, Li2017} examined the history of buoyancy-driven flows in Earth's mantle by approximating palaeo temperature distributions as well as variable parameters such as the thermal diffusivity.

Although there are problem-specific details in their methodologies, \cite{Lukyanenko2021, Subramanian2020, Kabanikhin2020, Liu2008, Li2017} use the same optimal control technique known as Direct Adjoint Looping (DAL). 
Using DAL, they minimize the cost functional
\begin{align}
    \hspace{-0.2cm}\J_f^X \hspace{-0.1cm} &\equiv \frac{1}{2} \Big\langle\Big| X'(x, t_f) \Big|^2\Big\rangle = \frac{1}{2} \Big\langle\Big| X(x, t_f) - \overline{X}(x, t_f) \Big|^2\Big\rangle, \label{costintro}
\end{align}
where the angled brackets $\langle \cdot \rangle$ denote a spatial integral over the problem domain and the trial solution $X$ has some deviation $X' \equiv X - \overline{X}$.
They construct sequences of trial solutions, such as $\{X_{n}(x, t)\}$, where each element $X_{n}$ is computed from a corresponding trial initial condition $X_{n}(x, 0)$.
Convergence to the target initial condition $\overline{X}(x, 0)$ is seldom guaranteed because the user provides an initial guess $X_{0}(x, 0)$.

Related 2D Navier--Stokes inverse problems have been studied analytically from several perspectives.
\cite{Imanuvilov1997} uses boundary conditions as control inputs (with a given initial condition) and verifies controllability of certain 2D fluid systems.
Later, the existence of solutions to a more demanding retrospective Navier--Stokes inverse problem with final overdetermination was established by \cite{Fan2009}.
In \cite{Fan2009}, the authors do not attempt to solve their inverse problem.
Instead they provide an analysis of the problem where a separable source term (as well as the initial condition) are left to be determined. 
Following this, the data-completion problem of determining an unknown viscosity was studied analytically by \cite{Fan2010}.

More recently, researchers extended these analytical studies of Navier--Stokes inverse problems by implementing data assimilation algorithms.
\cite{Du2013} studies an inverse problem governed by the parabolized Navier--Stokes (PNS) equation.
Following this, \cite{Law2012} studies data assimilation algorithms for non-deterministic Navier--Stokes inverse problems. 
They demonstrate several advantages of using filters for numerical weather prediction.
Yet more recently, \cite{Fan2020, FrerixThomas2021VDAw, Zhao2022, mowlavi_optimal_2023} used neural networks to solve analogous Navier--Stokes data-completion problems.
\hly{In particular, \cite{mowlavi_optimal_2023} compares DAL with a physics-informed neural network approach while solving inverse problems governed by the viscous Burgers equation as well as 2D incompressible Navier--Stokes.}

\hspace{-0.38cm}\hly{In this paper, we will also compare the performance of several iterative methods (such as DAL) by applying them to a retrospective Navier--Stokes inverse problem.}
Various Navier--Stokes inverse problems have been studied analytically and numerically using optimal control.
However, we will demonstrate that conventional optimal control methods are ill-suited for this class of inverse problems, even when a numerically-consistent dataset is available.

With DAL, we can minimize an arbitrary cost functional such as $\J_f^X$.
First, we must solve the forward problem on $t:0\to t_f$.
This forward problem is constrained by some PDE (such as Navier--Stokes) which we express generally as eqn \ref{genintro}.
At this point, we can evaluate $\J_f^X$ and compare the trial ($X(x, t_f)$) and target ($\overline{X}(x, t_f)$) final states.
Information from time $t=t_f$ is then propagated backward, to time $t=0$, by solving the linear adjoint problem $\mathcal{F}^{\dagger}[\chi, X]=0$ on $t:t_f\to 0$, where $\chi(x,t)$ denotes the adjoint variable.
Backward integration of the adjoint equations \colorbox{white}{allows us to compute} the gradient or functional derivative of $\J_f^X$ with respect to $X(x, 0)$. 
Using this gradient, we adjust our trial initial condition by some amplitude $\gamma$: $X_{n+1}(x, 0) = X_n(x, 0) + \gamma\chi_n(x, 0)$.


In practice, researchers often customize the DAL algorithm to improve performance in specific cases.
For example, \cite{Mannix2022discrete} used the conjugate-gradient method with a discrete adjoint formulation to optimize initial conditions subject to norm constraints.
For the PNS inverse problem, \cite{Du2013} combined Principle Orthogonal Decomposition (POD) with four-dimensional variational assimilation (4DVAR), an optimal control technique analogous to DAL.
In both examples, researchers reduced the computational cost of each DAL iteration by reformulating the adjoint while simultaneously reducing the required number of DAL iterations by implementing quasi-Newton minimization routines.
For this investigation, we take a different approach.
Instead of reformulating the adjoint or changing the way in which gradient information is used, we will demonstrate that DAL's shortcomings in the context of Navier--Stokes inversion can be mitigated by appending the adjoint system with additional advective terms.

The \colorbox{white}{additional advective} terms not present in DAL are present in two existing nonlinear backward integration methods: Simple Backward Integration (SBI) and the Quasi-Reversible Method (QRM). 
SBI and QRM are designed to invert the target final state by approximating the (ill-posed) integration of the original equations backward in time.
SBI accomplishes this by reversing the sign of the problematic diffusion term, i.e.~$a\partial_x^2 X \to -a\partial^2_x X$.
In contrast, QRM is a regularization framework which preserves the ill-posed diffusion term \cite{qrmsource}.
With QRM, we append a small hyperdiffusion term to the otherwise ill-posed constraint equation, i.e.~$a\partial_x^2 X \to a(\partial^2_x + \varepsilon\partial^4_x) X$ where $0<\varepsilon\ll 1$. 

We introduce these concepts as backward integration methods rather than iterative methods because their previous implementations were rarely iterated.
SBI (which has never been iterated) was introduced by \cite{Conrad2003} to infer the history of Earth's convective mantle in the viscinity of the southern African superswell.
Following this, \cite{Liu2008} and \cite{Li2017} used SBI to construct initial guesses $X_0(x,0)$, which they refined iteratively using DAL.
SBI was also examined by \cite{Fang2012} to dismiss a time-reversibility hypothesis on Subgrid-Scale (SGS) Large-Eddy Simulations (LES). 

\cite{darde2016} applied iterative QRM to ill-posed linear inverse problems and derives rigorous statements of convergence.
More recently, \cite{Nguyen2022} used iterative QRM to compute the unknown source in a nonlinear hyperbolic equation.
Using Carleman estimates, they demonstrate QRM's convergence to the correct source at an exponential rate.
While an analysis involving Carleman estimates is outside the scope of this paper, we refer to \cite{imanuvilov_2022}.


We compare DAL, SBI, and QRM by solving two retrospective inverse problems, respectively constrained by Korteweg--de Vries--Burgers (KdVB) and incompressible Navier--Stokes.
The novelty of our work lies in the \textit{iterative} application of SBI and QRM.
At the start of iteration $n$, we evolve some initial trial state $X_n(x,0)$ to its final state $X_n(x,t_f)$. 
Then, we initialize SBI and QRM using the final deviation $X'_n(x,t_f)$ in order to approximate the initial deviation $X'_n(x, 0)$.
Finally, assuming we know $X'_n(x, 0)$, we refine the trial state as $X_{n+1}(x, 0) := X_{n}(x, 0) - X'_n(x, 0)$ to conclude the iteration.
Equivalently, iterative SBI and QRM minimize the inaccessible objective functional 
\begin{align}
    \hspace{-0.2cm}\J_0^X \hspace{-0.1cm} &\equiv \frac{1}{2} \Big\langle\Big| X'(x, 0) \Big|^2\Big\rangle = \frac{1}{2} \Big\langle\Big| X(x, 0) - \overline{X}(x, 0) \Big|^2\Big\rangle \label{costoutro}
\end{align}
by repeatedly approximating its gradient $X'(x, 0)$.
$\J_0^X$ measures the trial state's error at the initial time, whereas $\J_f^X$ does so at the final time.

The remainder of this paper is organized as follows:
In Section \ref{secmethods} we provide a conceptual description of our iterative methods (DAL, SBI, and QRM).
In Section \ref{seckdvb} we illustrate these methods by applying them to a retrospective KdVB inverse problem. 
In Section \ref{secNS} we generalize these methods for multiple spatial dimensions, and compare their implementations using a 2D incompressible Navier--Stokes inverse problem.
In Section \ref{secCongeneralized}, we demonstrate that the adjoint equation which yields the gradient of $\J_0^X$ is ill-posed, nonlinear, and equivalent to the perturbed constraint eqn \ref{genintro}. 
In Section \ref{secCon} we summarize our results and discuss future applications for iterative SBI and QRM.

\section{Overview of Methods}\label{secmethods}

The Korteweg--de Vries--Burgers (KdVB) equation 
\begin{align}
  \mathcal{F}[\overline{u}(x,t)] \equiv \partial_t \overline{u} + \overline{u}\partial_x \overline{u} - a\partial_x^2 \overline{u} + b\partial_x^3 \overline{u} = 0 \label{kdvb}
\end{align}
provides a concise context for demonstrating each method (DAL, SBI, QRM).
Let $\overline{u}(x,t)$ denote to the target solution where $0\leq t\leq t_f$ in a 1D periodic domain $0 \leq x<L_x$.
The retrospective inverse problem is solved by recovering the target initial state $\overline{u}(x,0)$ where the target final state $\overline{u}(x,t_f)$ is given.
The general strategy has us construct trial solutions $u(x,t)$ satisfying eqn \ref{kdvb}, which we compute from trial initial states $u(x,0)$.
We measure a trial state's proximity from the target using the cost functionals $\Juo$ and $\Juf$, which are defined for an arbitrary variable $X$ in eqns \ref{costintro} and \ref{costoutro}.
Both $\Juo$ and $\Juf$ vary with respect to the trial initial state $u(x,0)$.
However, given that our knowledge of the target solution is limited to the final time $t_f$, it follows that we can only evaluate $\Juo$ if the target solution is known apriori.

\begin{figure}
  \centering
  \includegraphics[width=3.5in]{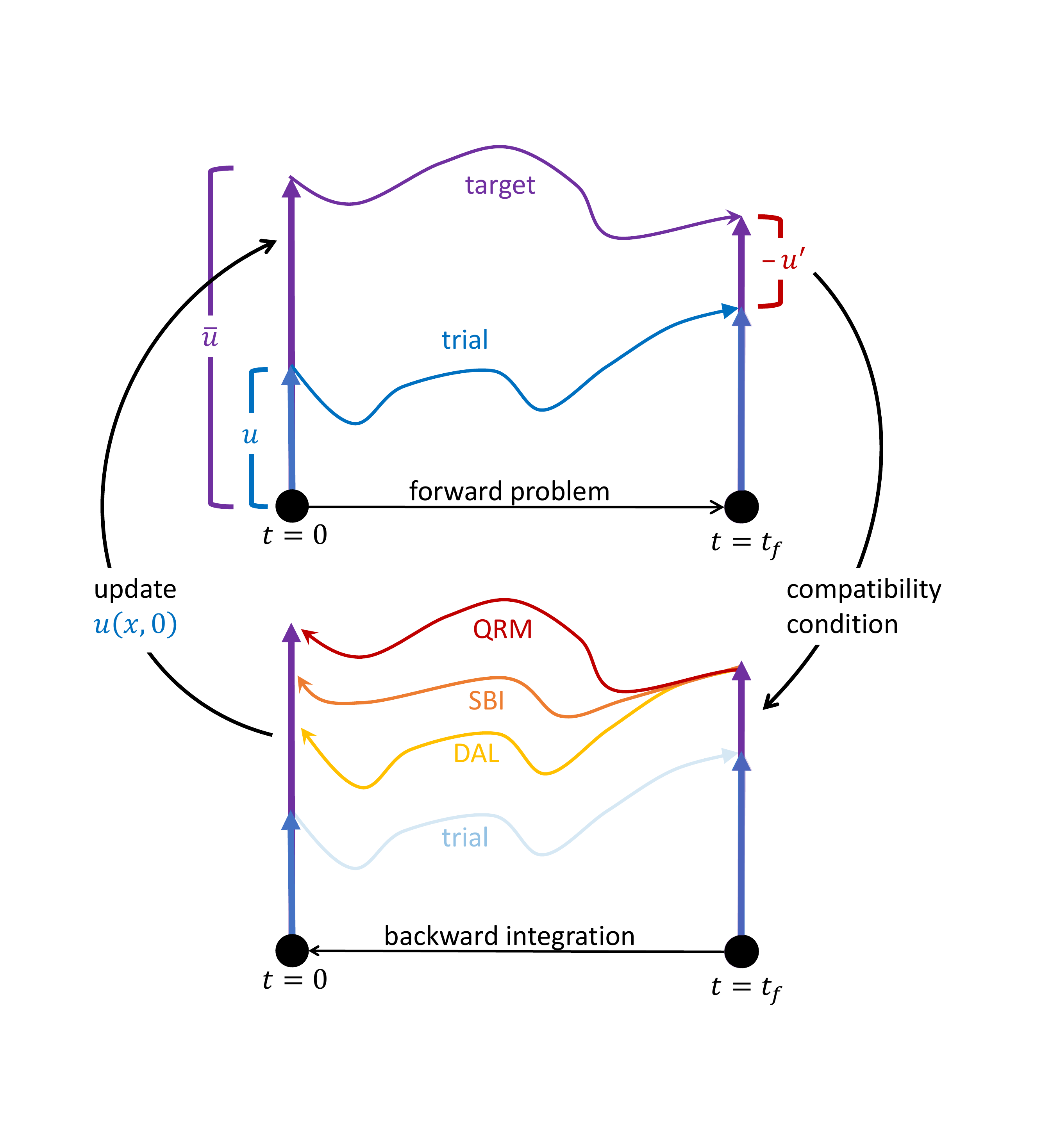}
  \caption{Conceptual illustration of iterative methods.
Purple: the target solution $\overline{u}(x, t)$ is known only at the final time $t_f$.
Blue: we evolve a trial state $u(x, 0)$ to $u(x, t_f)$ where we then initialize one of three backward integration systems using a single compatibility condition. 
Each backward integration gives a solution at $t=0$ which we use to update our trial initial condition. 
We seek the trial solution's deviation $u'=u-\overline{u}$ at the initial time $t=0$.
Yellow: Direct Adjoint Looping (DAL) is a linear backward integration.
For advective systems such as KdVB and Navier--Stokes, DAL advects the final deviation $u'(x,t_f)$ by the trial solution $u$. 
Orange: Simple Backward Integration (SBI) is a hybrid method, which supplements the linear DAL system with two additional advective terms.
Red: the Quasi-Reversible Method (QRM) aims to compute the deviation by introducing a small hyperdiffusion term with coefficient $\varepsilon$. 
The QRM backward integration is ill-posed when $\varepsilon=0$ and numerically unstable as $\varepsilon\to 0$.}
  \label{adj_diag}
\end{figure}

\subsection{Direct Adjoint Looping (DAL)}\label{secDAL}
In general, we can cast an inverse problem into the form of an optimization problem by defining and minimizing an appropriate cost functional \cite{Steiner2012}.
DAL provides an efficient avenue for computing a cost functional's gradient.
For KdVB, we minimize $\Juf$ by computing its gradient with respect to $u(x,0)$.
Once this gradient is known, we refine the trial initial state $u(x,0)$ by following the direction of steepest descent.

Using DAL, the gradient is computed as follows.
We define the Lagrangian $\L$
\begin{align}
  \L &\equiv \int_0^{t_f} \Big\langle \mu(x,t) \cdot \mathcal{F}[u(x,t)] \Big\rangle dt + \Juf
\end{align}
where $\mu$ is the adjoint variable corresponding to $u$.
$\L$'s variations with respect to $u$ and $\mu$ must disappear at the optimum $\overline{u}$, yielding
\begin{align}
  0<t<t_f: \quad  &0 =  \partial_t u + u\partial_x u - a\partial_x^2 u + b\partial_x^3 u \label{for}\\
  t=t_f:  \quad &\mu = -u' \label{fc}\\
  0<t<t_f: \quad &0 = \partial_t\mu + u\partial_x \mu + a\partial_x^2 \mu + b\partial_x^3 \mu \label{adj}.
\end{align}
The gradient of $\Juf$ resides in the manifold of solutions satisfying eqns \ref{for}, \ref{fc}, and \ref{adj}.
Figure \ref{adj_diag} illustrates how these equations are solved sequentially.
Given a trial initial condition, we start by solving the forward problem eqn \ref{for} from $t:0\to t_f$ (shown in blue). 
Next, we initialize the adjoint variable using the compatibility condition eqn \ref{fc}. 
Finally, we solve the adjoint eqn \ref{adj} backward in time $t:t_f\to 0$ (shown in yellow), yielding the desired gradient $-\mu(x,0)$. 
\hly{{The adjoint involves the trial solution $u$, so we must store data from the forward solve in memory. 
In practice this leads to redundant computations (due to memory constraints), offering another area for tangible improvement.}}

Once $\Juf$'s gradient is known, we use gradient descent with the popular Barzilai--Borwein method \cite{10.1093/imanum/8.1.141} to select step sizes. 
In addition to gradient descent, we also minimize $\Juf$ using the second-order sparse L-BFGS algorithm \cite{doi:10.1137/0916069} which utilizes the gradient's history over several iterations.
\hly{We select these two optimization routines for comparison after observing that they converge faster than alternatives. \footnote[2]{In addition to the results shown in section \ref{seckdvb}, we perform numerical experiments on the KdVB inverse problem using conjugate-gradient and line search routines (provided by scipy version 1.12.0).
In both cases we use the default parameters.
For the line search, we use the negative gradient for the descent direction.}}


\subsection{SBI and QRM Setup}\label{secsetup}
The KdVB retrospective inverse problem can be solved by finding the initial deviation $u'(x,0)$ of some trial state $u(x,0)$.
$u'=u - \overline{u}$ obeys a perturbation equation
\begin{align}
  0 = \partial_t u' - a\partial_x^2 u' + b\partial_x^3 u' + u\partial_x u' + u' \partial_x u - u'\partial_x u' \label{burgpert}
\end{align}
which is ill-posed over $t:t_f\to 0$.
We cannot compute $u'(x, t<t_f)$ numerically due to the intrinsic loss of information in diffusive systems.
DAL circumvents this obstacle by computing a gradient, whereas SBI and QRM approximate the deviation $u'$.
Let $\tmu$ denote our SBI and QRM approximations for $-u'$.

\subsection{Simple Backward Integration (SBI)}\label{secSBI}
SBI is one method for approximating the ill-posed backward integration, in which we reverse the sign of the problematic diffusion term, i.e.~$a\partial_x^2 \tmu \to -a\partial_x^2 \tmu$ such that
\begin{equation}
  0 = \partial_t \tmu + a\partial_x^2 \tmu + b\partial_x^3\tmu + u\partial_x\tmu + \tmu\partial_x u + \tmu\partial_x\tmu.\label{sbibi}
\end{equation} 
This backward integration is well-posed. 
Eqn \ref{sbibi} contains every term in eqn \ref{adj} (DAL) with two additional advective terms: $\tmu\partial_x u$ and $\tmu\partial_x\tmu$.

For the KdVB inverse problem, we first initialize eqn \ref{sbibi} with $\tmu(x, t_f) = -u'(x,t_f)$, just as with with DAL.
We then solve eqn \ref{sbibi} on $t:t_f\to 0$. 
Finally, SBI yields an approximation $\tmu(x,0)\approx-u'(x,0)$.
The SBI backward integration loop is illustrated in orange in Figure \ref{adj_diag}, where the additional SBI terms cause $\tmu$ to deviate from the linear DAL calculation of $\mu$.
After completing an SBI loop, we update the trial state $u_{n+1}(x,0) = u_n(x,0) + \tmu_n(x,0)$.
This is analogous to performing gradient descent with a fixed step size of unity.\footnote[3]{SBI and QRM do not generally return any sensible gradient, so the specified updates are not technically gradient descent. 
Attempting a constant step size of $\gamma\sim \mathcal{O}(1)$ with DAL is unstable for the direct minimization of $\Juf$. SBI can often be accelerated using a larger step size.}

In previous investigations, SBI was applied once to the target solution.
Our iterative implementation differs, because we apply SBI to the deviation equation.
Despite the inaccuracies introduced by reversing the sign of the diffusive term, we will demonstrate that iterative SBI is an effective algorithm.
SBI has already been implemented to develop initial guesses for retrospective inverse problems. 
We extend this work by comparing SBI with DAL rather than using them separately for different purposes.

\subsection{Quasi-Reversible Method (QRM)}\label{secQRM}
QRM deals with the ill-posed diffusion term by introducing a small, well-posed hyperdiffusion term: $a\partial_x^2 \tmu \to a(\partial_x^2 \tmu + \varepsilon \partial_x^4 \tmu)$. 
This method replaces eqn \ref{burgpert} by
\begin{equation}
  0 = \partial_t \tmu - a(\partial_x^2 \tmu + \varepsilon\partial_x^4 \tmu) + b\partial_x^3 \tmu + u\partial_x\tmu + \tmu\partial_x u + \tmu\partial_x\tmu.\label{qrmbi}
\end{equation} 
$0 < \varepsilon \ll 1$ is a free parameter chosen according to the numerical resolution as well as the solutions' spatial scales.
Notice how the ill-posed term remains, but the equation becomes well-posed because small-scale modes do not grow arbitrarily fast.

The QRM loop is carried out using the same procedure as SBI.
We initialize eqn \ref{qrmbi} with $\tmu(x, t_f) = -u'(x,t_f)$ and update the trial state $u_{n+1}(x,0) = u_n(x,0) + \tmu_n(x,0)$.
In Figure \ref{adj_diag}, we illustrate how QRM (red) recovers the unknown deviation precisely as $\varepsilon\to 0$.
However this limit is numerically unstable.

\section{KdVB Numerical Experiments}\label{seckdvb}
\begin{figure*}
  \centering
  \includegraphics[width=5in]{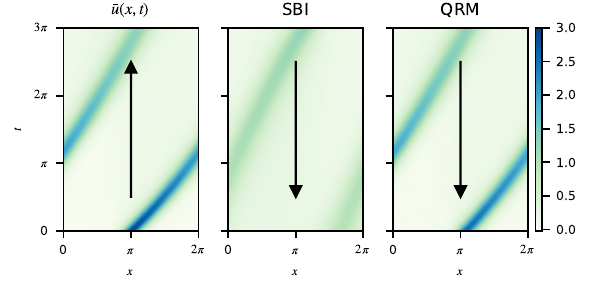}
  \caption{Target KdVB solution $\overline{u}(x,t)$ (left) plotted as a function of space $x$ and time $t$.
  We aim to recover the initial condition of a nonlinear wave which propagates right at an initial speed of one. 
The problem's spatial domain is 1D periodic. 
We include diffusion ($a=0.04$) such that its speed decreases as the wave propagates. 
Beginning at $t=3\pi$, SBI (center) evolves $\overline{u}(x,t_f)$ backward to $t=0$ by reversing the sign of the diffusive term.
The wave diffuses during the forward and backward integrations, such that its amplitude and speed of propagation decrease in both directions.
We also perform the QRM backward integration (right) on $t:3\pi\to 0$. 
Unlike SBI, QRM repopulates small-scale modes which are lost during forward integration. 
As the wave's peak regains its former amplitude, its speed of propagation increases mimicking $\overline{u}(x,t)$.
}
  \label{targy}
\end{figure*}

We implement each iterative method (DAL, SBI, QRM) to solve an inverse problem constrained by KdVB (eqn \ref{kdvb}).
The target solution $\overline{u}(x,t)$ (shown left in Figure \ref{targy}) consists of a dissipating nonlinear wave where $a=0.02$ and $b=0.04$. 
The spatial domain is 1D periodic $x\in [0, 2\pi]$ and we solve eqn \ref{kdvb} over $t:0\to3\pi$.
The target initial condition
\begin{align}
  \overline{u}(x,0) &= 3\cosh^{-2} \left( \frac{x-\pi}{2\sqrt{b}} \right) \label{kdvbic},
\end{align}
would propagate rightwards as a stable soliton with a constant speed of one if diffusion were not present (i.e.~if $a=0$ in eqn \ref{kdvb}).
This retrospective problem is nontrivial, as the final state cannot be evolved backwards in time via conventional timestepping algorithms.

The target and trial solutions along with each backward integration are carried out using the \texttt{Dedalus} open-source pseudospectral python framework \cite{Burns2020}. 
We represent $u,\;\mu,$ and $\tmu$ as vectors of 128 real Fourier modes. 
While computing $u(x, t)$, we store the solution vector at each timestep.
Linear (diffusive and dispersive) terms are treated implicitly whereas the nonlinear advective term is evaluated explicitly by multiplying $u$'s grid data on a 3/2 dealiased grid. 
The same holds when we perform backward integrations, where terms such as $\mu\partial_x u$ must be treated explicitly to avoid dense matrix operations at every timestep.
\hly{Compared to DAL, the SBI and QRM backward integration equations contain more of these terms. 
Consequently, each SBI and QRM iteration requires approximately $30\%$ more computation time.}
For every solve we use a 2nd-order Runge--Kutta scheme with a fixed timestep $\Delta t=0.01$.
\hly{We verify the adjoint solver's accuracy by comparing its output to a gradient obtained via finite-difference.
Our code for this paper is available at \url{https://github.com/liamoconnor9/adjop}}

We initialize each iterative method using an initial guess $u^0(x,0)=0$.
(See Appendix \ref{secAppKdVBSBI} for the same comparison beginning with an SBI initial guess.)
The first SBI iteration is shown in the center panel of Figure \ref{targy}.
Diffusion of the nonlinear wave causes its amplitude and speed of propagation to decrease such that the wave's peak does not return to its original position.
The first QRM iteration (shown right in Figure \ref{targy}) reconstructs the wave by inverting the advection alongside diffusion.
In this case, the wave's peak returns to its approximate initial position.
For QRM we use $\varepsilon=0.01$ whereas SBI involves no free parameter.
\begin{figure*}
  \centering
  \includegraphics[width=5.0in]{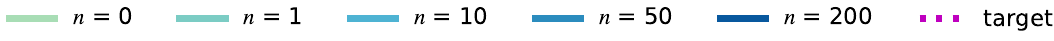}
  \includegraphics[width=5.0in]{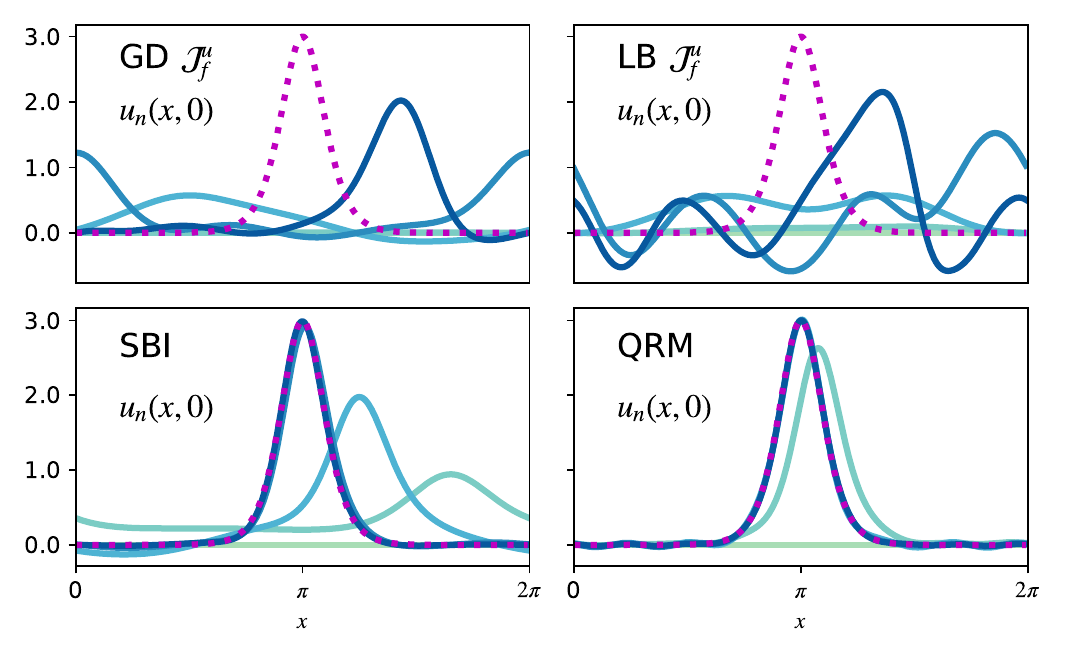}
  \caption{Evolution of the KdVB trial initial conditions $u_n(x,0)$ using an initial guess $u_0(x,0)=0$ (shown in green).
  We compare two DAL runs (top row) with SBI (bottom left) and QRM (bottom right).
  (Top left) gradient descent approaches the target initial state gradually.
  (Top right) DAL minimization of $\Juf$ using L-BFGS approaches the target state more rapidly, but this method introduces undesired medium-scale features.
  SBI captures the target state faster and more directly than DAL.
  QRM resembles the target state immediately $(n=1)$. 
  However, this method introduces low-amplitude small-scale errors which do not subside at $n=200$.
  }
  \label{kdvevo}
\end{figure*}

The top row of Figure \ref{kdvevo} illustrates the evolution of DAL trial initial states $u_n(x,0)$.
We minimize $\Juf$ using gradient descent (top left) and L-BFGS (top right).
For advective systems such as KdVB, the amplitude of each feature is proportional to its speed of propagation.
Consequently, gradient descent gives a sequence of initial states which increase in amplitude while shifting leftwards.
$u_{10}(x,0)$ contains a shallow peak centered near $x=\pi/2$.
$u_{50}(x,0)$ is centered near $x=0$ with approximately twice the amplitude of $u_{10}(x,0)$.
Similarly, $u_{200}(x,0)$ is centered near $x=3\pi/2$ with approximately triple the amplitude of $u_{10}(x,0)$.
Their respective final states ($u_{10}(x,3\pi)$, $u_{50}(x,3\pi)$, and $u_{200}(x,3\pi)$) all contain peaks which approximately align with the target's ($\overline{u}(x,3\pi)$).

Rather than following the direction of steepest descent, L-BFGS anticipates changes in the gradient by constructing a sparse Hessian representation.
Although this quasi-Newton method recovers the target initial state better than gradient descent at $n=200$, L-BFGS still does not yield a quality approximation of $\overline{u}(x,0)$.
This case introduces additional wave features in $u_{10}(x,0)$, $u_{50}(x,0)$, and $u_{200}(x,0)$ which do not appear in the target.

To locate the target state, we introduce additional advective terms in the backward integration.
The bottom row of Figure \ref{kdvevo} illustrates the evolution of trial initial states $u_n(x,0)$ refined via SBI (bottom left) and QRM (bottom right) over 200 iterations.
SBI does not effectively approximate the initial deviation $u'_n(x,0)$ (as shown for $n=0$ in Figure \ref{targy}), nor does it follow the direction of steepest descent of $\Juf$.
This algorithm smoothly guides the initial states' peaks directly toward the target.
In this case, $u_{50}(x,0)$ is indistinguishable from $\overline{u}(x,0)$.
QRM loosely captures the target's structure in a single iteration by approximating the initial deviation $u'_n(x,0)$.
In this case, $u_{10}(x,0)$ is nearly indistinguishable from $\overline{u}(x,0)$.

After one iteration, the SBI and QRM initial states' peaks are centered rightwards of the target feature (see $u_1(x,0)$ in Figure \ref{kdvevo}, bottom left and right panels). 
These nonlinear backward integrations couple the deviation's amplitude with the resulting direction of our update for $u_n(x,0)$.
SBI refines each trial initial state by advecting the final deviation backward in time as it continues to diffuse. 
This decreases the speed of backward advection such that each update does not return to the position of the initial deviation.
This is illustrated by $u_{10}(x,0)$ and $u_{50}(x,0)$ in the bottom left panel of Figure \ref{kdvevo}.
Their peaks shift leftward, approaching the target with each iteration.
When the trial final state intercepts the target, the deviation shrinks and SBI's backward diffusion has less influence on the direction of each subsequent update.
QRM refines each trial state by advecting the final deviation while simultaneously inverting diffusion.
The hyperdiffusion term slows backward advection to a lesser extent than SBI, yielding smaller rightward misalignments. 
These updates approximate the initial deviation as it shrinks with each iteration, such that iterative QRM behaves like a one-shot method (e.g.~\cite{Nabi2022}).

\begin{figure}
  \centering
  \includegraphics[width=3.5in]{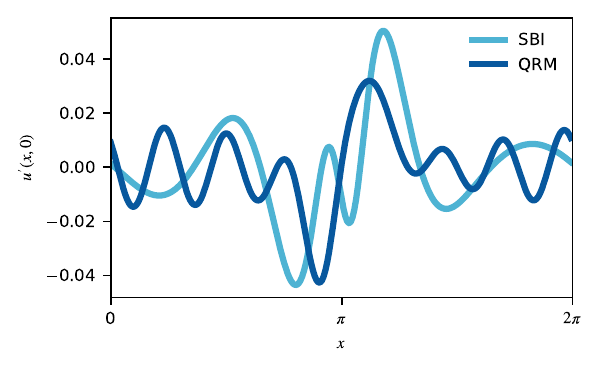}
  \caption{Deviation of SBI and QRM trial initial conditions, plotted as a function of $x$ at iteration $n=200$. 
  Both curves have prominent peaks which overlap near $x=\pi$. 
  The SBI deviation has a slightly larger magnitude while the QRM deviation oscillates at a particular wavenumber.}
  \label{kdvb_diff}
\end{figure}

\hspace{-0.45cm}\hly{For QRM and SBI, we plot the deviation $u'(x,0) = u_{200}(x,0) - \overline{u}(x,0)$ at iteration $n=200$ in Figure \ref{kdvb_diff}.
Both curves have relatively large features coinciding with the target initial condition's peak at $x=\pi$.
The SBI deviation has a larger maximum magnitude while the QRM deviation oscillates at a particular wavenumber.
When we increase $\varepsilon$, these oscillations subside, but it requires more iterations to obtain a comparable result.
We select $\varepsilon = 0.01$ to illustrate the presence of these oscillations, while still achieving the desired outcome in a small number of iterations.
}

\begin{figure}
  \centering
  \includegraphics[width=3.5in]{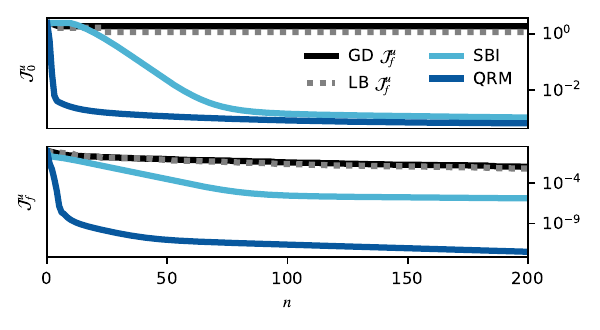}
  \caption{Initial errors ($\Juo$, top) and final errors ($\Juf$, bottom) for the 1D KdVB inverse problem plotted as a function of iteration $n$.
  We compare DAL, SBI, and QRM. 
  DAL performs slightly better when paired with the second-order sparse minimization routine L-BFGS (LB $\Juf$), whereas gradient descent (GD $\Juf$) performs the worst.
  Our iterative applications of SBI and QRM minimize $\Juo$ and $\Juf$ more effectively than DAL. 
  QRM minimizes both $\Juo$ and $\Juf$ more rapidly than SBI.}
  \label{kdvb_optimize}
\end{figure}

Errors from each optimization study are plotted in Figure \ref{kdvb_optimize}.
DAL does not effectively minimize $\Juo$ or $\Juf$.
In similar investigations, L-BFGS has received significant attention thanks to its DAL-compatible features \cite{Li2017}. 
It converges faster than gradient descent which is also reflected in our experiment.
However, the SBI and QRM iterative methods both outperform DAL by several orders of magnitude.
QRM minimizes both $\Juo$ and $\Juf$ more rapidly than SBI, as shown in Figure \ref{kdvb_optimize}.
At $n\approx 100$, SBI stalls and QRM approaches floating-point precision in terms of $\Juf$.
We attribute this gap to an energy discrepancy between the SBI and QRM updates.
For QRM applied to a highly diffusive system, the energy $\langle|\tmu(x,t)|^2\rangle$ increases during backward integration (provided $\varepsilon$ is sufficiently small) such that $\langle|\tmu(x,0)|^2\rangle \sim \langle|u'(x,0)|^2\rangle$.
This is illustrated in the first QRM iteration (Figure \ref{targy}, right).
In contrast, SBI diffuses during backward integration (Figure \ref{targy}, center) such that $\langle|\tmu(x,0)|^2\rangle < \langle|u'(x,0)|^2\rangle$.
We can eliminate this energy discrepancy and often accelerate SBI by increasing the step sizes of the updates (e.g.~$u_{n+1}(x,0)=u_{n}(x,0)+1.1\mu_{n}(x,0)$).

\section{2D Navier--Stokes Numerical Experiments}\label{secNS}

Consider 2D flow with spatial coordinate $\vec{x} = x\vec{\hat{x}} + y\vec{\hat{y}}$. 
The velocity $\vec{\overline{u}}(\vec{x},t)$ obeys the incompressible Navier--Stokes equation
\begin{align}
  &\partial_t \vec{\overline{u}} + \vec{\overline{u}}\cdot\grad \vec{\overline{u}} + \grad \overline{p} = \nu\grad^2\vec{\overline{u}} \quad \textrm{and} \quad \grad\cdot\vec{\overline{u}} = 0 \label{2dns}
\end{align}
  {where $\overline{p}$ is the pressure and $\rm{Re} \equiv \nu^{-1} = 50,000$. 
  The domain is doubly-periodic with $x\in[0,1]$ and $y\in[-1,1]$, such that its boundary conditions are given by $\vec{\overline{u}}|_{x = 0} = \vec{\overline{u}}|_{x = 1}$ and $\vec{\overline{u}}|_{y = -1} = \vec{\overline{u}}|_{y = 1}$. 
  
  For our inverse problem, we use divergence-cleaning to construct the target initial condition $\vec{\overline{u}}(\vec{x}, 0)$.
  First, consider the following velocity field which has a nonzero divergence}
\begin{align}
    \vec{\overline{v}}(\vec{x})\cdot\vec{\hat{x}} &= \frac{1}{2}\bigg[ \tanh\bigg(10\Big(y-\frac{1}{2}\Big)\bigg) - \tanh\bigg(10\Big(y+\frac{1}{2}\Big)\bigg) \bigg]\; \nonumber ;\\
\vec{\overline{v}}(\vec{x})\cdot\vec{\hat{y}} &= \frac{1}{10}\sin\big( 2\pi x \big)\Bigg[\exp\bigg( -100\Big(y - \frac{1}{2}\Big)^2 \bigg) \\
&\qquad\qquad\qquad\qquad - \exp\bigg( -100\Big(y + \frac{1}{2}\Big)^2 \bigg)  \Bigg].\nonumber
\end{align}
Next, we perform divergence cleaning on $\vec{\overline{v}}$ by solving the boundary value problem $\grad^2 p_v + \grad \cdot \vec{\overline{v}} =0$. 
The divergence-free target initial condition is then given by $\vec{\overline{u}}(\vec{x},0) = \vec{\overline{v}}(\vec{x}) + \grad p_v(\vec{x})$.

For our inverse problem, we run a target simulation by evolving $\vec{\overline{u}}(\vec{x}, 0)$ according to eqn \ref{2dns} until $t_f=20$.
The target initial condition has two thin layers of positive and negative vorticity concentrated near $y=1/2$ and $y=-1/2$.
Advection causes these vortex sheets to wind counter-clockwise and clockwise while viscosity smooths its small-scales.

Given the final velocity field $\vec{\overline{u}}(\vec{x}, t_f)$, we develop trial solutions $\vec{u}(\vec{x},t)$ (also obeying eqn \ref{2dns}) which are meant to approach the target solution $\vec{\overline{u}}(\vec{x},t)$.
All PDE solves (except the GD $\JUf$ double-resolution study) are performed on a 128 by 256 grid of real Fourier modes using the \texttt{Dedalus} pseudospectral framework \cite{Burns2020}. 
We timestep with $\Delta t=0.002$ using a second-order Runge--Kutta scheme. 
Nonlinear operations are evaluated on a 3/2 dealiased grid.
We plot the upper half the domain (unit square) because the solutions have an approximate symmetry about $y=0$.
The target solution's vorticity evolution is illustrated in the top row of Figure \ref{snapshots200}.

\subsection{DAL}
\label{NSdal}
Let $\vec{\mu}$ and $\Pi$ be the adjoint variables respectively corresponding to $\vec{u}$ and $p$.
The Navier--Stokes adjoint equations are then given by
\begin{align}
  0 &= \grad\cdot\vec{\mu} \label{nsadj} \\
  \vec{0} &= \partial_t \vec{\mu} + \vec{u}\cdot\grad\vec{\mu} \highlightmath{- \vec{\mu}\cdot(\grad\vec{u})^T} + \grad\Pi + \nu\grad^2\vec{\mu}. \nonumber
\end{align}
\cite{Kerswell2014} provides a detailed derivation of eqn \ref{nsadj}.
Using DAL, we minimize two objective functionals with associated compatibility conditions
\begin{align}
  \JUf &\equiv \frac{1}{2}\left\langle|| \vec{u'}(\vec{x}, t_f)||^2\right\rangle   \rightarrow  \vec{\mu}(\vec{x}, t_f)=-\vec{u'}(\vec{x}, t_f); \\
  \Jwf &\equiv \frac{1}{2}\left\langle\left( \omega'(\vec{x}, t_f)\right)^2\right\rangle \rightarrow  \vec{\mu}(\vec{x}, t_f)=\grad^{\perp}\omega'(\vec{x}, t_f) \label{Jw}.
\end{align}
The skew-gradient $\grad^{\perp}\equiv (-\partial_y, \; \partial_x)$ and the vorticity $\omega \equiv \grad^{\perp}\cdot\vec{u}$. 
As with KdVB, a prime denotes the deviation of a trial variable from its target, e.g.~$\omega'\equiv\omega-\overline{\omega}$ and so on.
$\JUf$ quantifies the trial states' proximity to the target at the final time $t_f$ in terms of velocity whereas $\Jwf$ does so using vorticity.
We implement the gradient descent (GD) and L-BFGS optimization routines to minimize these two cost functionals.
\hly{We verify the accuracy of the adjoint solver by using the finite-difference approximation to reproduce the gradient at low resolution.}

\subsection{SBI and QRM}
\label{NSsbiqrm}
Our iterative procedures for SBI and QRM are analogous to those described in the previous section. 
The ill-posed backward integration of $\vec{u'}$ obeys the nonlinear perturbation equation
\begin{align}
  0 &= \grad\cdot\vec{u'} \label{nspert}\\
  \hspace{-2cm}\vec{0} &= \partial_t \vec{u'} + \vec{u}\cdot\grad \vec{u'} + \vec{u'}\cdot\grad \vec{u} - \vec{u'}\cdot\grad\vec{u'} +\grad p' - \nu\grad^2 \vec{u'}.  \nonumber
\intertext{
Next we substitute the approximations $-\vec{\tmu}$ for $\vec{u'}$ and $-\tilde{\Pi}$ for $p'$.
For SBI, we reverse the sign of the viscous term $\nu\grad^2 \vec{\tmu}$. 
Its corresponding approximation for eqn \ref{nspert} is given by}
  0 &= \grad\cdot\vec{\tmu} \\
  \hspace{-2cm}\vec{0} &= \partial_t \vec{\tmu} + \vec{u}\cdot\grad \vec{\tmu} + \vec{\tmu}\cdot\grad \vec{u} + \vec{\tmu}\cdot\grad\vec{\tmu} +\grad \tilde{\Pi} + \nu\grad^2 \vec{\tmu}\nonumber
\intertext{For QRM, we introduce a small hyperdiffusion term, approximating eqn \ref{nspert} as}
  0 &= \grad\cdot\vec{\tmu} \\
  \hspace{-2cm}\vec{0} &= \partial_t \vec{\tmu} + \vec{u}\cdot\grad \vec{\tmu} + \vec{\tmu}\cdot\grad \vec{u} + \vec{\tmu}\cdot\grad\vec{\tmu} +\grad \tilde{\Pi} \nonumber \\
  &\qquad\qquad\qquad\qquad\qquad\qquad\qquad\qquad - \nu(\grad^2 + \varepsilon\grad^4) \vec{\tmu}.\nonumber
\end{align}
\hly{For this problem, we set $\varepsilon=0.001$, because the oscillatory errors described in Section \ref{seckdvb} trigger a numerical instability when $\varepsilon \lesssim 0.001$.}

\begin{figure}
  \centering
  \includegraphics[width=3.5in]{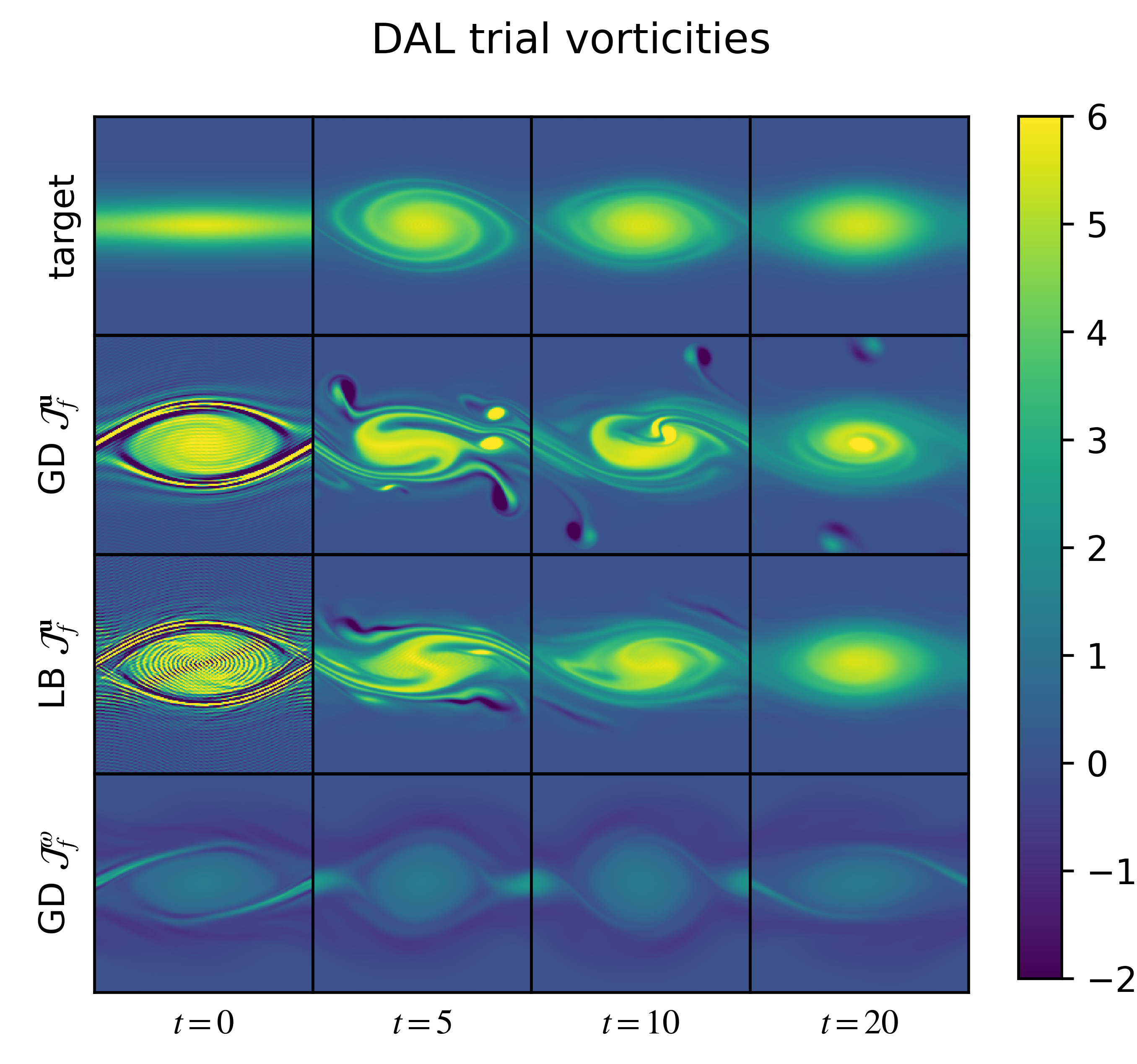}
  \caption{Target (top row) and trial vorticity snapshots in the upper half of the domain at simulation times $t=0,5,10,20$. 
  The second, third, and bottom row trial solutions are obtained by applying DAL over 200 iterations.
  GD $\JUf$ (second row) uses gradient descent to minimize $\JUf$. 
  For this case, we use double resolution in space and time (256 by 512 modes, $\Delta t=0.001$) to ensure that each simulation is resolved.
  Despite this extra resolution, we observe symmetry breaking when $t\geq 5$ due to numerical errors.
  LB $\JUf$ (third row) uses L-BFGS to minimize $\JUf$ with the default resolution.
  Although the initial condition is under-resolved in this case, we still recover a final state which resembles the target.
  GD $\Jwf$ (bottom row) uses gradient descent to minimize $\Jwf$. 
  The DAL trial initial conditions (bottom three rows) have small-scale banded features with high concentrations of positive vorticity.
  Their final states have varying degrees of resemblance to the target at $t=20$.
  }
  \label{snapshots200}
\end{figure}

\subsection{Results}
We initialize each iterative method using an initial guess $\vec{u}^0(\vec{x},0)=\vec{0}$.
For DAL, we minimize the velocity error using gradient descent (GD $\JUf$) and L-BFGS (LB $\JUf$).
We run GD $\JUf$ at double resolution to ensure that the trial state's evolution is well-resolved.
We also minimize the vorticity error via gradient descent (GD $\Jwf$).
L-BFGS minimization of $\Jwf$ frequently gives unstable trial initial conditions which cannot be evolved without additional resolution.
Figure \ref{snapshots200} consists of vorticity snapshots belonging to the target solution and trial solutions at $t=0,5,10,20$.
The trial solutions in these cases are obtained using DAL until iteration $n=200$.
GD $\JUf$ (second row) contains highly concentrated bands of positive vorticity encompassing a larger, lower-amplitude vortex.
Although this case is resolved, we observe symmetry breaking when $t\gg 0$.
This trial solution adequately captures the target at $t=20$, but not as well as LB $\JUf$.
LB $\JUf$ (third row) is clearly under-resolved at $t=0$, with the same highly-concentrated features as GD $\JUf$ along with small-scale ringing throughout the domain.
These ringing features subside after a brief interval ($t\geq 5$), after which, the trial solution approaches the target as $t\to 20$.
GD $\Jwf$ (bottom row) is well-resolved, but its vortex's amplitude is appreciably lower than that of the target.
In this case, we still observe band-like regions of concentrated vorticity encompassing a larger low-amplitude vortex.
The DAL trial initial conditions (illustrated in the left column, lower three rows) do not share similar structures with the target (top left).
Despite this, advection sweeps the concentrated vorticity bands into a single coherent vortex after 20 advective times.
This process illustrates ill-conditioning of the Navier--Stokes inverse problem.


\begin{figure}
  \centering
  \includegraphics[width=3.5in]{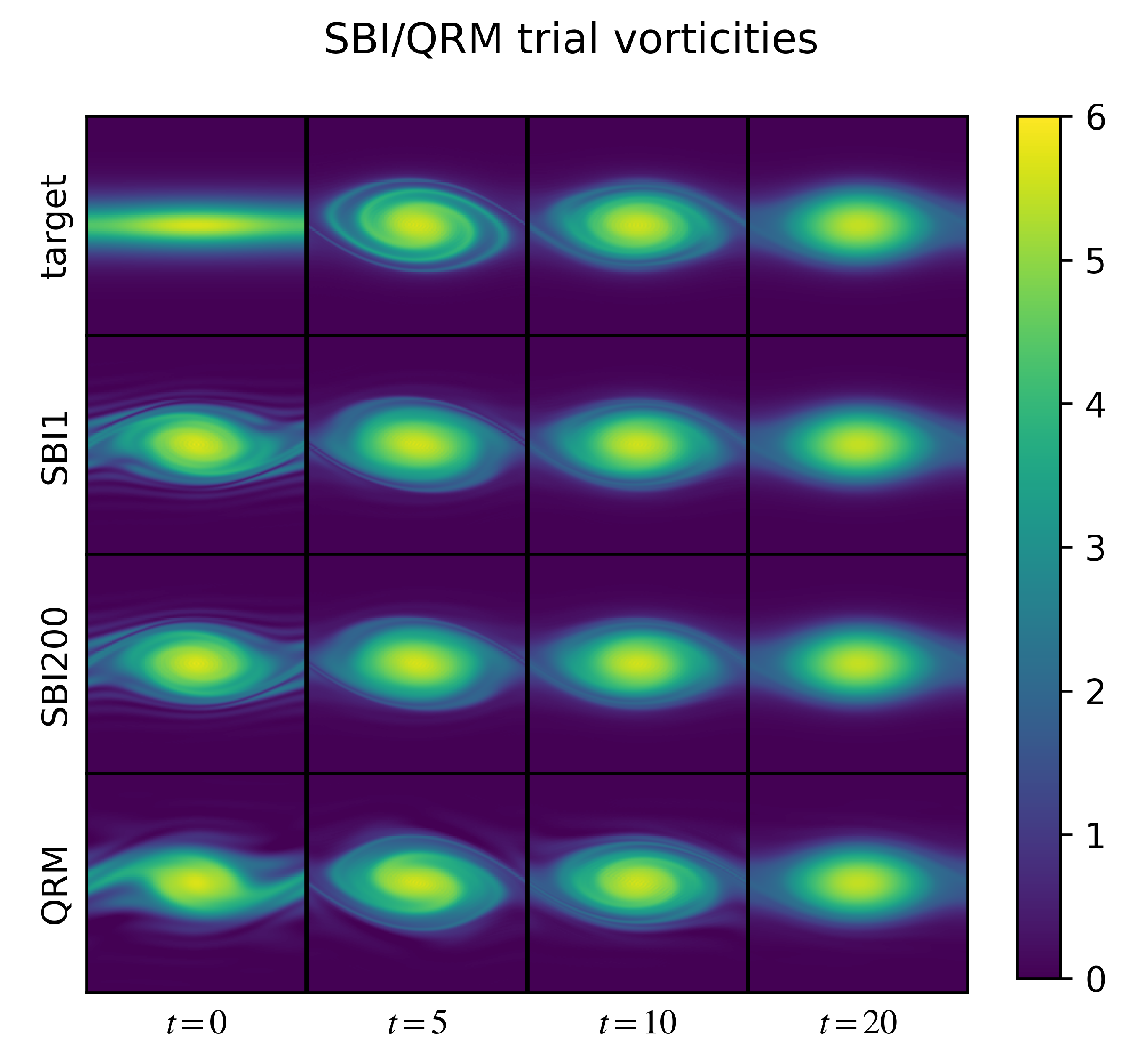}
  \caption{Same as Figure \ref{snapshots200} using the SBI and QRM iterative methods.
  Despite the ill-conditioning which hinders DAL in Figure \ref{snapshots200}, SBI and QRM both recover the target initial condition (top left) to a greater extent.
  SBI1 and SBI200 (middle two rows) denote SBI trial states after 1 and 200 iterations respectively.
  The QRM initial state (bottom left) loosely resembles the target initial condition at $n=200$.}
  \label{snapshots200_sbi}
\end{figure}

Figure \ref{snapshots200_sbi} illustrates the analogous trial vorticity snapshots using SBI and QRM.
In the top row, we reproduce the target.
The second row (SBI1) denotes the trial state after a single SBI iteration.
Its initial state resembles the target's better than any of the DAL runs in Figure \ref{snapshots200}.
Its final state is nearly indistinguishable from the target.
The third row (SBI200) denotes the SBI trial solution after 200 iterations.
The bottom row (QRM) denotes the trial solution after 200 QRM iterations.
This case shares the most resemblance with the target throughout the time domain.
At $t=0$, QRM has curved arm-like features which mimic the target's initially confined shear layer.
At $t\geq 5$, this trial solution resembles the target.
In general, SBI and QRM capture the target state better than DAL without introducing the undesired features shown in Figure \ref{snapshots200}.

\begin{figure}
  \centering
  \includegraphics[width=3.5in]{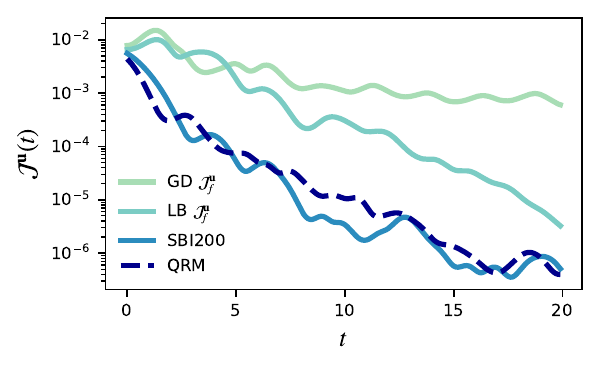}
  \caption{Velocity error $\Ju(t)\equiv\frac{1}{2}\big\langle |\vec{u}(\vec{x},t)-\vec{\overline{u}}(\vec{x},t)|^2\big\rangle$ vs.~time of Navier--Stokes trial solutions at iteration $n=200$. GD $\JUf$ and LB $\JUf$ denote DAL minimization of $\JUf$ via gradient descent and L-BFGS respectively. SBI200 and QRM denote the trial solutions refined by SBI and QRM respectively.
  Every trial solution's error descreases nonmonotonically with respect to time.
  QRM has the smallest error at the initial and final times, though SBI200 has less error over large intervals in the time domain.}
  \label{traces}
\end{figure}

Figure \ref{traces} shows the time-dependent velocity error $\J^{\vec{u}}(t)\equiv\frac{1}{2}\langle| \vec{u}(\vec{x},t) - \vec{\overline{u}}(\vec{x},t)|^2\rangle$ for the trial solutions GD $\JUf$, LB $\JUf$, SBI, and QRM at iteration $n=200$.
Due to the inverse-problem's ill-conditioning, all trial states begin with substantially larger errors than their respective final states.
These errors decrease nonmonotonically on $t:0\to 20$.
The DAL trial solutions' (GD $\JUf$ and LB $\JUf$) errors increase on $0<t<2$ as their small-scale vortex bands (illustrated in the left column of Figure \ref{snapshots200}) rapidly disappear. 
LB $\JUf$ performs significantly better than GD $\JUf$ throughout most of the time domain, especially near the final state.
In contrast, the velocity errors of SBI and QRM decrease immediately after initialization.
Throughout the time domain, SBI and QRM perform better than either DAL case.

\begin{figure}
  \centering
  \includegraphics[width=3.5in]{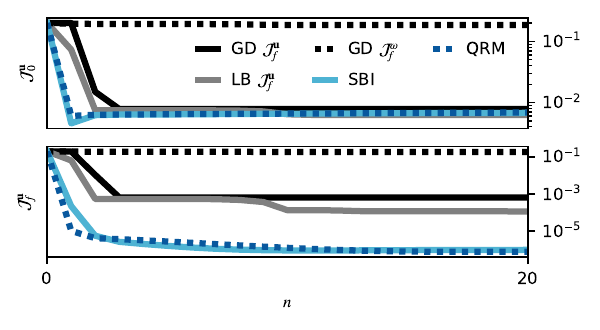}
  \caption{Initial (top, $\JUo$) and final (bottom, $\JUf$) velocity errors for the 2D Navier--Stokes inverse problem over 20 iterations starting with an initial guess $\vec{u}^0(\vec{x}, 0)=\vec{0}$. 
  We compare how Direct Adjoint Looping (DAL), Simple Backward Integration (SBI), and the Quasi-Reversible Method (QRM) minimize these errors. 
  For DAL we compute the gradients of $\JUf$ and $\Jwf$ (shown black and grey).
  Using the gradient of $\JUf$, we implement gradient descent and L-BFGS (solid black and solid grey respectively). 
  $\JUf$ decreases rapidly when SBI and QRM are used. 
  Their respective performances are comparable in the first 20 iterations.
  }
  \label{shearstudyq}
\end{figure}

Figure \ref{shearstudyq} plots the initial ($\JUo$, top) and final ($\JUf$, bottom) velocity errors vs.~iteration $n$ for each iterative method until $n=20$.
Every method's performance is comparable in terms of $\JUo$ except GD $\Jwf$ which does not appreciably decrease this error.
In terms of $\JUf$, SBI (light blue solid) and QRM (dark blue dotted) both outperform every DAL case by several orders of magnitude.  

\begin{figure}
  \includegraphics[width=3.5in]{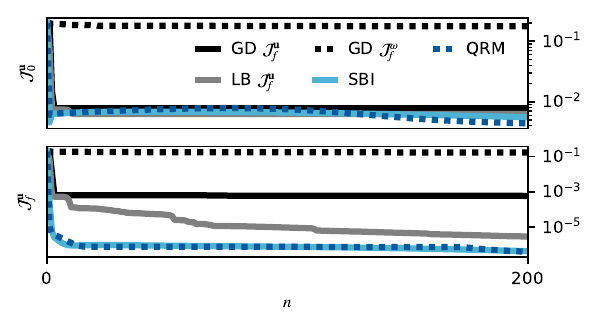}
  \caption{Same as Figure \ref{shearstudyq} extended over 200 loops. 
  DAL minimization of $\Jwf$ (dotted black) is ineffective in terms of both $\JUo$ (top) and $\JUf$ (bottom).
  After the initial 20 iterations, every other method decreases $\JUo$ by less than an order of magnitude.
  DAL minimization of $\JUf$ using L-BFGS (grey, LB $\JUf$) is the most effective DAL case in terms of both errors.
  At $n=200$, LB $\JUf$ does not minimize $\JUf$ to the same extent as SBI and QRM do in very few iterations.
  Figures \ref{snapshots200} and \ref{snapshots200_sbi} illustrate their corresponding trial vorticities, where it is clear that LB $\JUf$ is under-resolved and does not approximate the target initial state.
  }
  \label{shearstudyq2}
\end{figure}

Figure \ref{shearstudyq2} extends Figure \ref{shearstudyq} to iteration $n=200$.
Both DAL gradient descent algorithms (GD $\JUf$ and GD $\Jwf$) stall after $n=20$, whereas LB $\JUf$ continues to minimize the final error $\JUf$.
The relative success of LB $\JUf$ indicates that the gradient of $\JUf$ has large curvature.
SBI and QRM continue to outperform all DAL runs, with QRM decreasing $\JUo$ slightly more.
With QRM, the errors decrease nonmonotonically whereas SBI exhibits a smooth decline.

Whereas previous work used SBI and QRM to perform a single loop, we demonstrate in Figures \ref{shearstudyq} and \ref{shearstudyq2} that these methods can be iterated.
SBI and QRM both decrease the errors by more than an order of magnitude in a single iteration.
For $n>150$ in Figure \ref{shearstudyq2}, SBI and QRM continue to gradually minimize $\JUf$.
This prolonged minimization of $\JUf$ using iterative SBI is a surprising result of our study.
In terms of $\JUo$, every method fails to recover the target state due to ill-conditioning of this inverse problem.
QRM surpasses the other methods in this respect, as illustrated in Figure \ref{shearstudyq2}.

\section{Derivation of Ill-Posed Nonlinear Adjoint Equation}
\label{secCongeneralized}
Consider the KdVB inverse problem from Sections \ref{secmethods} and \ref{seckdvb}.
Throughout this investigation, we assert that SBI and QRM perform better than DAL because they approximate the initial deviation rather than computing a gradient.
In fact, the gradient of $\Juo$ taken with respect to $u(x,0)$ is the unknown deviation $u'(x,0)=u(x,0)-\overline{u}(x,0)$.
$\Juo$ measures the squared distance between the trial and target initial states.
Its levelsets form hyperspheres centered around $\overline{u}(x,0)$ and its paths of steepest descent are radial (pointing directly at $\overline{u}(x,0)$).
The gradient of $\Juo$ embeds global information, whereas the gradient of $\Juf$ gives local information in the space of trial initial conditions.

Using optimal control, we will demonstrate that the ill-posed backward problem (eqn \ref{burgpert}) is equivalent to the adjoint which minimizes $\Juo$.
First we multiply eqn \ref{burgpert} by $u'$
and integrate over the space-time domain, yielding $\Juo$ in terms of $u'$, $u$, and $\Juf$.
\begin{align}
  \Juo = \Juf - \int_0^{t_f}\langle au'\partial_x^2 u' - u' \partial_x [uu'] \rangle dt \label{energy}. 
\end{align}
Eqn \ref{energy} amounts to an energy equation for the deviation between our target and trial states. 
  Notice how terms which are conservative of the $L_2$ norm cancel ($u'\partial_x u'$ and $b\partial_x^3 u'$) whereas terms with non-zero energy flux ($\partial_x [uu']$ and $a\partial_x^2 u'$) persist.
  Using eqn \ref{energy} we construct a new Lagrangian 
\begin{align}
  \hat{\L} &= \int_0^{t_f} \langle \hat{\mu}(x,t) \cdot \mathcal{F}[u(x,t)] \rangle dt + \Juo \\
  &= \int_0^{t_f} \Big\langle \hat{\mu}\big[ \partial_t u + u\partial_x u - a\partial_x^2 u + b\partial_x^3 u \big] \nonumber \\
  &\qquad\qquad\qquad\qquad - au'\partial_x^2 u' + u' \partial_x [uu' ] \Big\rangle dt + \Juf. \nonumber
\end{align}
with a corresponding adjoint variable $\hat{\mu}$. 
Taking the variation of $\hat{\L}$ with respect to $\hat{\mu}(x,t)$ still returns the original constraint eqn \ref{kdvb} (on $0<t<t_f$) and compatibility condition eqn \ref{fc} (at $t=t_f$). 
However, varying $\hat{\L}$ with respect to $u(x,t)$ along $0<t<t_f$ gives a new adjoint equation 
\begin{align}
0 &= -\partial_t \hat{\mu} - u\partial_x \hat{\mu} - a\partial_x^2 \hat{\mu} - b\partial_x^3 \hat{\mu} \nonumber \\
&\qquad - 2a\partial_x^2 u' + u'\partial_x u - u'\partial_x u' \label{wadj}.
\end{align}

Finally, by summing eqns \ref{burgpert} and \ref{wadj}, it is evident that $\hat{\mu}(x,t) + u'(x,t) = 0$ not just at $t=t_f$, but throughout the time domain. 
We substitute $\hat{\mu}\to -u'$ in $\hat{\L}$'s adjoint, yielding
\begin{align}
  0 = \partial_t \hat{\mu} - a\partial_x^2 \hat{\mu} + b\partial_x^3 + u\partial_x \hat{\mu} + \hat{\mu}\partial_x u - \hat{\mu}\partial_x \hat{\mu} \label{illp}.
\end{align}
This new adjoint (equivalent to eqn \ref{burgpert}) is ill-posed and effectively nonlinear.
The same demonstration can be applied for any retrospective inverse problem with an advective constraint, including the Navier--Stokes inverse problem outlined in Section \ref{secNS}.
If eqn \ref{illp} could be solved, we would obtain the exact deviation in our trial initial condition by performing a single adjoint loop. 
For example, the non-diffusive Korteweg--de Vries equation and the inviscid Euler equation admit well-posed nonlinear adjoints as these equations are time-reversible.
However when $a>0$ (diffusive forward problem), eqn \ref{illp} cannot be solved backwards using conventional numerical solvers.
The ill-posed backward integration problem motivates the SBI and QRM modifications.
As $\varepsilon\to 0$, QRM backward integration approaches the ill-posed nonlinear adjoint which minimizes $\Juo$.
\hly{From this perspective, QRM is a regularized form of DAL.
SBI backward integration combines the linear adjoint which minimizes $\Juf$ with the additional terms appearing in the ill-posed nonlinear adjoint.}

Given that the SBI and QRM backward integrations return approximations for the gradient of $\Juo$, it follows that we can repeatedly apply the approximate gradients to refine our trial state.
Taking the dynamical systems perspective, we can treat these iterative methods as discrete maps where the target state is a fixed point.
Our results demonstrate that these fixed points are generally attracting for advection--diffusion systems.
Potential difficulties might arize in future investigations due to limit-cycles or even strange attractors.
We have simultaneously highlighted an important limitation of using gradients to optimize initial conditions: gradients can have large curvatures, even when the trial state is in close proximity to the extremum (see Appendices \ref{secAppKdVBSBI} and \ref{secAppNSBI}).
By including additional terms in our backward integrations, we relax the gradient's sensitive dependence on the trial initial condition by replacing the gradient with a smoother discrete map.
Higher-order algorithms for finding fixed points could also be paired with SBI and QRM to further accelerate convergence.

\section{Discussion and Conclusions}
\label{secCon}
\subsection{Summary of Results}
\label{secConSummary}
In this investigation, we extend several previous studies involving Direct Adjoint Looping (DAL), Simple Backward Integration (SBI), and the Quasi-Reversible Method (QRM) by applying them to a pair of retrospective inverse problems. 
DAL has been used previously to solve inverse problems and optimize simulation inputs, as it allows us to compute the gradient of an arbitrary functional.
This gradient can be used to refine the trial initial condition toward a local extremum.
However, the number of iterations required to approximate said extremum depends on the user's initial guess.

The SBI and QRM methods are designed to approximate the trial solution's deviation rather than taking a gradient.
Using the approximate deviation at the initial state, we update the trial initial condition using a ``step size'' of unity.
In practice, we cannot compute the initial deviation because the necessary backward integration is ill-posed.
SBI and QRM provide well-posed approximations for the ill-posed backward integration.

SBI has been used to make coarse approximations and generate initial guesses for further refinement via DAL.
We extend this use by developing an iterative algorithm which acts on the trial state's deviation.
With this method, we evolve the trial state's final deviation $u'(x,t_f)$ backwards in time using a modified constraint equation.
This modified constraint equation is well-posed because we reverse the sign of the diffusive term.

Iterative QRM has previously been implented.
With this backward integration, we preserve the ill-posed diffusion term by introducing a small hyperdiffusion term to relax the numerical instability.
We must also tune this hyperdiffusion term's coefficient $\varepsilon$ according to the problem's dynamics and numerical resolution.

Our first inverse problem is constrained by the Korteweg--de Vries--Burgers (KdVB) equation.
The target solution consists of a nonlinear wave whose advective velocity decays due to diffusion.
We demonstrate that the gradient of $\J^u_f$, obtained via DAL or otherwise, is not an efficient means of recovering the target initial state.
DAL performs poorly when coupled with gradient descent as well as the L-BFGS optimization routine.
When following the direction of steepest descent in the space of trial initial conditions, we accumulate undesired wave features.
This is due to high curvature in the path of steepest descent.
QRM captures the target final state almost immediately while SBI does so more gradually.

The second inverse problem is constrained by the 2D incompressible Navier--Stokes equation.
The target solution is a symmetric Kelvin--Helmholtz vortex with $\rm{Re}\equiv\nu^{-1}=50,000$.
We evolve the target state over 20 advective time units, then use the final state to recover its associated initial condition.
The problem domain is doubly-periodic which likely reduces the ill-conditioning of our retrospective inverse problem.
We apply each iterative method until iteration $n=200$.
Here we utilize the flexibility of DAL by minimizing the velocity error $\JUf$ as well the vorticity error $\Jwf$.
DAL minimization of $\Jwf$ performs poorly as illustrated in Figures \ref{snapshots200}, \ref{shearstudyq} and \ref{shearstudyq2}.
Using L-BFGS to minimize $\JUf$, we produce an initial condition whose final state resembles the target after 200 iterations (Figure \ref{snapshots200}).
This L-BFGS case successfully minimizes $\JUf$, as illustrated in Figure \ref{shearstudyq2}.
However, the trial initial condition is under-resolved because L-BFGS accumulates small-scale features.
This highlights severe ill-conditioning in the Navier--Stokes inverse problem, as we are able to minimize the final error without approximating the target initial condition.
We implement gradient descent with double resolution to demonstrate this more clearly.
This gradient descent case does not approximate the target final state as well as L-BFGS, even though the simulations are well-resolved.

Just as with KdVB, SBI and QRM outperform DAL by every metric in the Navier--Stokes inverse problem.
Although every method decreases $\JUo$ with a similar trend (Figures \ref{shearstudyq} and \ref{shearstudyq2}), SBI and QRM (Figure \ref{snapshots200_sbi}) approximate the target initial state better than DAL (Figure \ref{snapshots200}) at $n=200$.
The methods' respective behaviors differ in close proximity to their target, as illustrated in Figure \ref{shearstudyq2}.
Specifically, QRM exhibits nonmonotonicity in both $\JUo$ and $\JUf$ whereas SBI decreases these quantities with each iteration.


\subsection{Future Work}
\label{secConfuture}
Optimal control via DAL is the most common approach for retrospective inverse problems.
However, SBI and QRM are more effective than DAL when applied to the 2D Navier--Stokes inverse problem.
\hly{In future work, we will implement SBI and QRM for 3D Navier--Stokes inverse problems.
Chaos and the direct energy cascade (as opposed to the inverse energy cascade in 2D) will likely present additional challenges.
We will also use SBI and QRM to solve other optimal control problems, such as the mantle convection inverse problems studied by \cite{Liu2008} and \cite{Li2017}.}
Numerical weather prediction (NWP) could also be enhanced by implementing these nonlinear methods.
SBI and QRM are well-suited for this application because NWP models are increasingly nonlinear.
Predictions are limited by our ability to incorporate real-time observations in a reasonable number of iterations \cite{Bonavita2018}.

Another relevant study is \cite{Ambrose2015}, which computes time-periodic solutions of vortex sheets with surface tension.
Using DAL, they minimize a cost functional which resembles $\Juf$.
Although they are not afforded a target final state, we can interpret their adjoint loop as a dynamic inversion where the target state changes with each iteration.
Conceptually, any properly-constrained optimal control problem has a corresponding retrospective inverse problem.
The compatibility condition may differ (e.g.~\cite{Ambrose2015} initializes their adjoint with the deviation between their trial initial and final states), but in general the adjoint evolution equations remain the same.
Given that our SBI and QRM loops tend toward the mutual extremum of $\Juo$ and $\Juf$, it is reasonable to suggest that these methods might also have attracting extrema when seeking periodic solutions.
Efficient algorithms for computing time-periodic solutions would have direct applications in studying cyclic dynamos and anticipating nonlinear resonances in engineering systems such as turbines.

\section*{Acknowledgments}
The authors thank Emma Kaufman, Benjamin Hyatt, Whitney Powers, Ilaria Fontana, and Adrian Fraser for their valuable suggestions.
We thank the \texttt{Dedalus} development team. 
L.O., D.L., K.A., K.J.B., J.S.O., and B.P.B. are supported in part by NASA HTMS grant 80NSSC20K1280. D.L., K.J.B., J.S.O., and B.P.B. are supported in part by NASA OSTFL grant 80NSSC22K1738. 
L.O. is supported by the National Science Foundation Graduate Research Fellowship under Grant No. (DGE-2234667).
Computations were conducted with support by the NASA High End Computing Program through the NASA Advanced Supercomputing (NAS) Division at Ames Research Center on Pleiades with allocation GID s2276.

\appendix\label{secAppendix}
\section{KdVB Inverse Problem Objectives using SBI Initial Guess}\label{secAppKdVBSBI}

\begin{figure*}
  \centering
  \includegraphics[width=5.0in]{legend_kdvb.pdf} \\
  \includegraphics[width=5.0in]{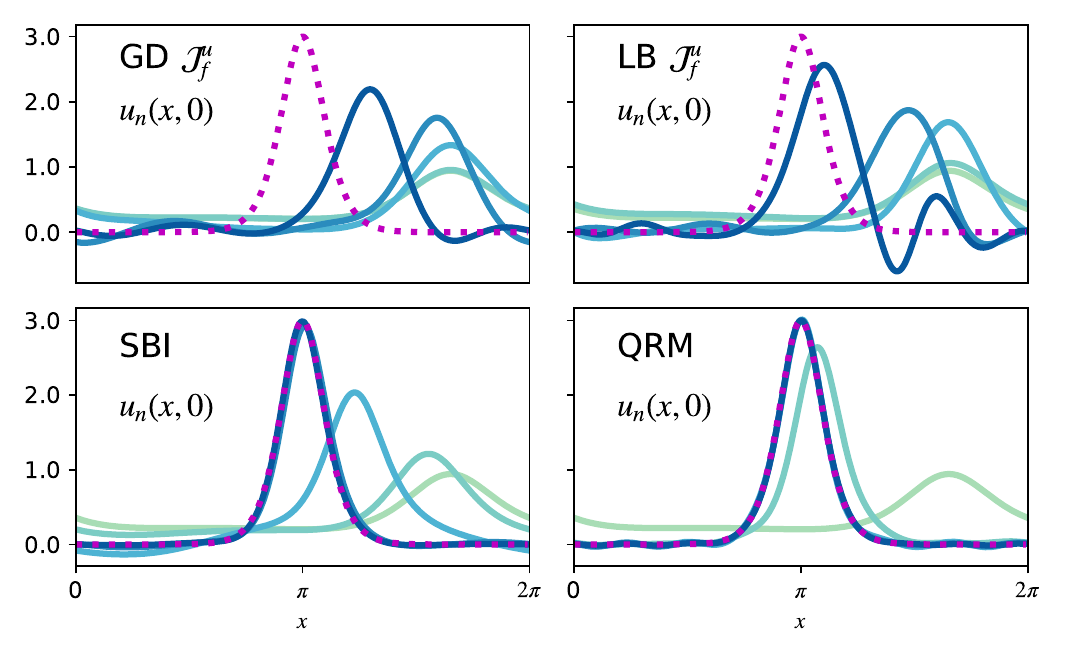}
  \caption{Same is Figure \ref{kdvevo} using the SBI initial guess (shown in green).
  }
  \label{skdvevo}
\end{figure*}

For the KdVB inverse problem described in Section \ref{seckdvb}, we initialize $u^0(x,0)$ using an initial guess constructed via SBI.
This SBI initial guess is computed by evolving $\overline{u}(x,t_f)$ backwards using a modified constraint equation (eqn \ref{kdvb} with the diffusive term's sign reversed). 
For DAL, this implementation mimics \cite{Li2017,Liu2008} who demonstrated that their SBI initial guess converges more rapidly compared to a neutral guess of zero.

Using the SBI initial guess does improve our DAL approximations of $\overline{u}(x,0)$, as shown in the top row of Figure \ref{skdvevo}.
L-BFGS (top right) continues to outperform gradient descent (top left), though both cases have significant deviations from $\overline{u}(x,0)$.
SBI and QRM both outperform the DAL cases, converging in approximately the same number of iterations as in Section \ref{seckdvb}.

\begin{figure}
  \centering
  \includegraphics[width=3.5in]{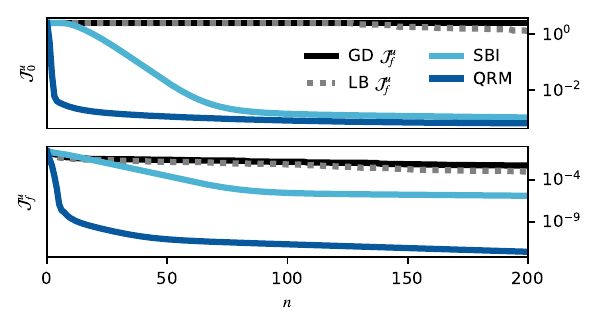}
  \caption{Same is Figure \ref{kdvb_optimize} using initial guess $u^0(x, 0)$ generated via SBI.}
  \label{skdvb_optimize}
\end{figure}

Figure \ref{skdvb_optimize} plots the initial (top) and final (bottom) errors for the KdVB inverse problem using the SBI initial guess.
LB $\Juf$ (shown in grey) performs appreciably better than in Figure \ref{kdvb_optimize}.
However SBI and QRM both surpass DAL, with QRM performing best by several orders of magnitude in terms of $\Juf$.

\section{Navier--Stokes Inverse Problem Objectives using SBI Initial Guess}\label{secAppNSBI}

\begin{figure}
  \centering
  \includegraphics[width=3.5in]{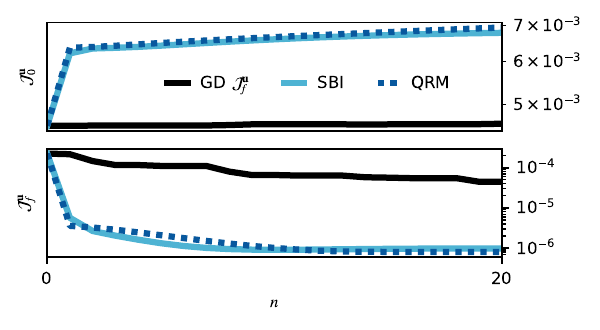}
  \caption{Same as Figure \ref{shearstudyq} using initial guess $\vec{u}^0(\vec{x}, 0)$ generated via SBI.}
  \label{shearstudys}
\end{figure}

\begin{figure}
  \centering
  \includegraphics[width=3.5in]{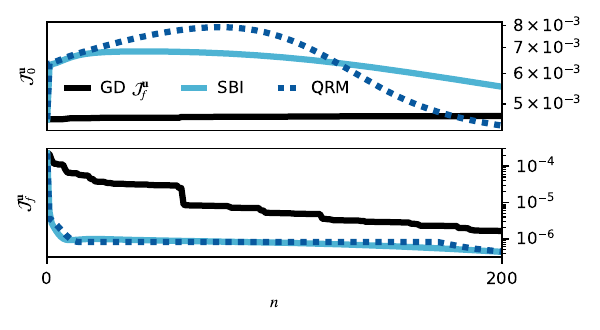}
  \caption{Same as Figure \ref{shearstudyq2} using initial guess $\vec{u}^0(\vec{x}, 0)$ generated via SBI.}
  \label{shearstudys2}
\end{figure}

For the Navier--Stokes inverse problem described in Section \ref{secNS}, we compare each method (DAL, SBI, and QRM) starting with an initial guess generated via SBI. 
This case mimics \cite{Li2017,Liu2008}, who began by rewinding their target state from $t:t_f\to 0$ using SBI. 
They subsequently refined their SBI initial guess by using DAL to minimize an error functional (analogous to DAL $\JUf$, shown in black). 
Figures \ref{shearstudys} and \ref{shearstudys2} plot the initial ($\JUo$, top) and final ($\JUf$, bottom) velocity errors as functions of iteration $n$.
Figure \ref{shearstudys} shows the first 20 iterations while Figure \ref{shearstudys2} terminates at $n=200$.
In the black curve (GD $\JUf$), we use DAL with gradient descent to minimize $\JUf$.

First, we confirm that using an SBI initial guess with DAL does approach the target final state faster than in Section \ref{secNS}. 
Next we implement SBI and QRM as before. 
Just as with the previous initial guess $\vec{u}^0(\vec{x}, 0)=\vec{0}$, we find that SBI and QRM outperform DAL, albeit by a narrower margin (ranging from less than one to two orders of magnitude in terms of $\JUf$).
Crucially, SBI and QRM both outperform DAL in very few iterations, suggesting that the efficacy of these methods is somewhat robust to the initial guess.
Every iterative method increases the initial error $\JUo$ (top) compared to the SBI initial guess, except for QRM which eventually decreases $\JUo$ below its original value near $n\approx 190$ (shown in dotted dark blue).
Iterative SBI appears to follow a similar trend at a slower pace, whereas GD $\JUf$ shows no such trend.
This behavior with GD $\JUf$ indicates that our Navier--Stokes inverse problem is highly non-convex, even in close proximity to the extremum.
In this case, L-BFGS minimization of $\JUf$ gives unstable trial initial conditions.
We obtain similar results by initializing each iterative method using the QRM trial initial state at $n=200$.

\section*{References}
\bibliography{bib.bib}

\end{document}